\documentclass{ws-ijmpd}

\begin{document}

\markboth{A. Bernui, \, M. J. Rebou\c{c}as, \, and \, A. F. F. Teixeira}
{A Study of Gaussianity in CMB band maps}

%
\catchline{}{}{}{}{}
%

\title{\uppercase{\bf A Study of Gaussianity in CMB band maps}}

\author{ARMANDO BERNUI$^{\,1}$}

\address{$^{1}$Instituto de Ci\^encias Exatas, \\
Universidade Federal de Itajub\'a, 37500-903, Itajub\'a \,--\, MG, Brazil
\\
abernui@gmail.com}

\author{MARCELO J. REBOU\c{C}AS$^{\,2}$ \ and \ ANTONIO F. F. TEIXEIRA$^{\,2}$}


\address{$^{2}$Centro Brasileiro de Pesquisas F\'\i sicas, \\
22290-180, \, Rio de Janeiro -- RJ, Brazil \\
reboucas@cbpf.br \,;\, teixeira@cbpf.br}

\maketitle

\begin{abstract}
The detection of non-Gaussianity in the CMB data would rule out a number
of inflationary models. A null detection of non-Gaussianity, instead, would
exclude alternative models for the early universe.
Thus, a detection or non-detection of primordial non-Gaussianity in the CMB
data is crucial to discriminate among inflationary models, and to test alternative
scenarios.
However, there are various non-cosmological sources of non-Gaussianity.
This makes important to employ different indicators in order to detect distinct
forms of non-Gaussianity in CMB data.
Recently, we proposed two new indicators to measure deviation from
Gaussianity on large angular scales, and used them to study the Gaussianity
of the \emph{raw} band WMAP maps with and without the \emph{KQ75} mask.
Here we extend this work by using these indicators to perform similar analyses
of deviation from Gaussianity of the \emph{foreground-reduced}  Q, V,
and W band maps.
We show that there is a significant deviation from Gaussianity in the considered
full-sky maps, which is reduced to a level consistent with Gaussianity when the
\emph{KQ75} mask is employed.
\end{abstract}

\keywords{Cosmic microwave background; non-Gaussianity;
physics of the early universe.}

\section{Introduction}

Cosmic Microwave Background (CMB) data from the Wilkinson Microwave
Anisotropy Probe (WMAP)\cite{WMAP1} is under rigorous scrutiny
for the possible deviations from Gaussianity\cite{non-Gauss}\cdash\cite{BRb}
and for statistical isotropy\cite{RELATED} in the CMB temperature field.
A detection or non-detection of primordial non-Gaussianity in the CMB data is
crucial to discriminate among inflationary models, and to test alternative
scenarios.\cite{Bartolo}
However, since non-cosmological effects can produce non-Gaussianity
in the CMB data, the extraction of a possible \emph{primordial}
non-Gaussianity is not a simple enterprise.
Besides, one does not expect that any single statistical estimator
can be sensitive to all possible forms of non-Gaussianity.
It is therefore important to test CMB data for deviations from Gaussianity
by using different statistical tools, since they can potentially provide
information about multiple forms of non-Gaussianity that may be present
in CMB data.
In a previous work\cite{BR} we introduced two indicators, based on
skewness and kurtosis of CMB data, and used them to study the deviation
from Gaussianity in the WMAP \emph{raw} band (uncleaned) maps.
In the present work, we complement this study of WMAP five-year data by
performing a similar analysis but now using the \emph{foreground-reduced}
Q, V, and W band maps with and without the \emph{KQ75} mask.
We show that there is a significant deviation from Gaussianity in these
\emph{foreground-reduced} full-sky band maps, which is reduced to a level
that is consistent with Gaussianity when the \emph{KQ75} mask is used.

\section{Non-Gaussian indicators}

In this section we  briefly present how to construct our non-Gaussinity indicators.
For a detailed discussion of these indicators we refer the readers to
Ref.~\refcite{BR} and Ref.~\refcite{BRb}.

Consider a discrete set of points $\{j=1, \ldots ,N_\mathrm{c} \}$ uniformly distributed
on the sphere $S^2$ as the center of spherical caps of a given aperture $\gamma$,
and calculate for each cap $j\,$ the real numbers 
\begin{eqnarray}
S_j   \equiv \frac{1}{N_{\mbox{\footnotesize p}} \,\sigma^3_j }
\sum_{i=1}^{N_{\mbox{\footnotesize p}}} \left(\, T_i\, - \overline{T}_j \,\right)^3 \; ,
\hspace{1cm}
K_j  \equiv \frac{1}{N_{\mbox{\footnotesize p}} \,\sigma^4_j }
\sum_{i=1}^{N_{\mbox{\footnotesize p}}} \left(\,  T_i\, - \overline{T}_j \,\right)^4 - 3 \;,
\nonumber
\end{eqnarray}
where $N_{\mbox{\footnotesize p}}$ is the number of pixels in the $j^{\,\rm{th}}$ cap,
$T_i$ is the temperature at the $i^{\,\rm{th}}$ pixel, $\overline{T}_j$ is the CMB mean
temperature on the $j^{\,\rm{th}}$ cap, and $\sigma_j$ is the standard deviation for each $j$.
The sets of values $\{S_j, j=1,\ldots,N_\mathrm{c}\}$ and $\{K_j, j=1,\ldots,N_\mathrm{c}\}$
can then be viewed as measures of the non-Gaussianity in the directions $(\theta_j, \phi_j)$
of the center of the $j^{\,\rm{th}}$ cap.
Patching together the $\{S_j\}$ and $\{K_j\}$ values, we obtain the skewness
$S = S(\theta,\phi)$ and kurtosis $K = K(\theta,\phi)$ functions defined on $S^2$.
Throughout this work the Mollweide projection of the functions $S = S(\theta,\phi)$
and $K = K(\theta,\phi)$ are named, respectively,  $S-$map and $K-$map.
Clearly, one can expand each of these functions in their spherical harmonics
and calculate their angular power spectrum.
For the skewness function $S = S(\theta,\phi)= \{S_j, \,j=1,\ldots,N_\mathrm{c}\}$,
for example, one has
\begin{eqnarray}
S (\theta,\phi) = \sum_{\ell=0}^\infty \sum_{m=-\ell}^{\ell}
b_{\ell m} \,Y_{\ell m} (\theta,\phi)
\,\, \Longrightarrow \,\,
S_{\ell} = \frac{1}{2\ell+1} \sum_m |b_{\ell m}|^2 \; , \nonumber
\end{eqnarray}
where $S_{\ell}$ is the corresponding angular power spectrum.
Similar expressions hold for the $K = K(\theta,\phi)$.

To obtain quantitative information for each multipole component $\ell$
about the non-Gaussianity at large-angles (low $\ell$) from the $S$ and $K$ maps
(obtained from the band maps) we calculate the power spectra $S_{\ell}$ and $K_{\ell}$.
Then, to estimate the statistical significance of each multipole,
we compare $S_{\ell}$ and $K_{\ell}$ to the mean multipole values $\overline{S}_{\ell}$
and $\overline{K}_{\ell}$ calculated by averaging over $1\,000$ power
spectra of $S$ and $K$ maps obtained from Monte Carlo (MC) Gaussian CMB
maps.\footnote{For details on both the MC Gaussian CMB maps
and the calculation of the mean power spectra $\overline{S}_{\ell}$ and
$\overline{K}_{\ell}$ see Refs.~\refcite{BR}
and~\refcite{BRb}.}

\section{Results and Conclusions}

In this section we report the results of our Gaussianity analysis performed
with the indicators discussed in the previous section.

To minimize the statistical noise, in the calculations
of $S$ and $K$ maps from the foreground-reduced Q, V, W input maps, 
we have scanned the celestial sphere with spherical caps of aperture  
$\gamma = 90^{\circ}$, whose centers are located at $12\,288$ points
homogeneously distributed on the celestial sphere.
The input maps we have used in our analyses have the resolution
HEALPix parameter $N_{\mbox{\footnotesize side}}= 256$,\cite{Gorski-et-al-2005} 
which corresponds to a partition of the celestial sphere into $786\,432$ pixels.
The calculations of the $S$ and $K$ maps by scanning the CMB
masked maps sometimes include caps whose center is within or close to
the \emph{KQ75} masked region. In these cases, the calculations are made with
a smaller number of pixels $N_{\mbox{\footnotesize p}}$. Finally, we note 
that examples of $S$ and $K$ maps can be found in Refs.~\refcite{BR} and 
\refcite{BRb}. 

\begin{figure}
\mbox{\hspace{-1.0cm}
\epsfig{file=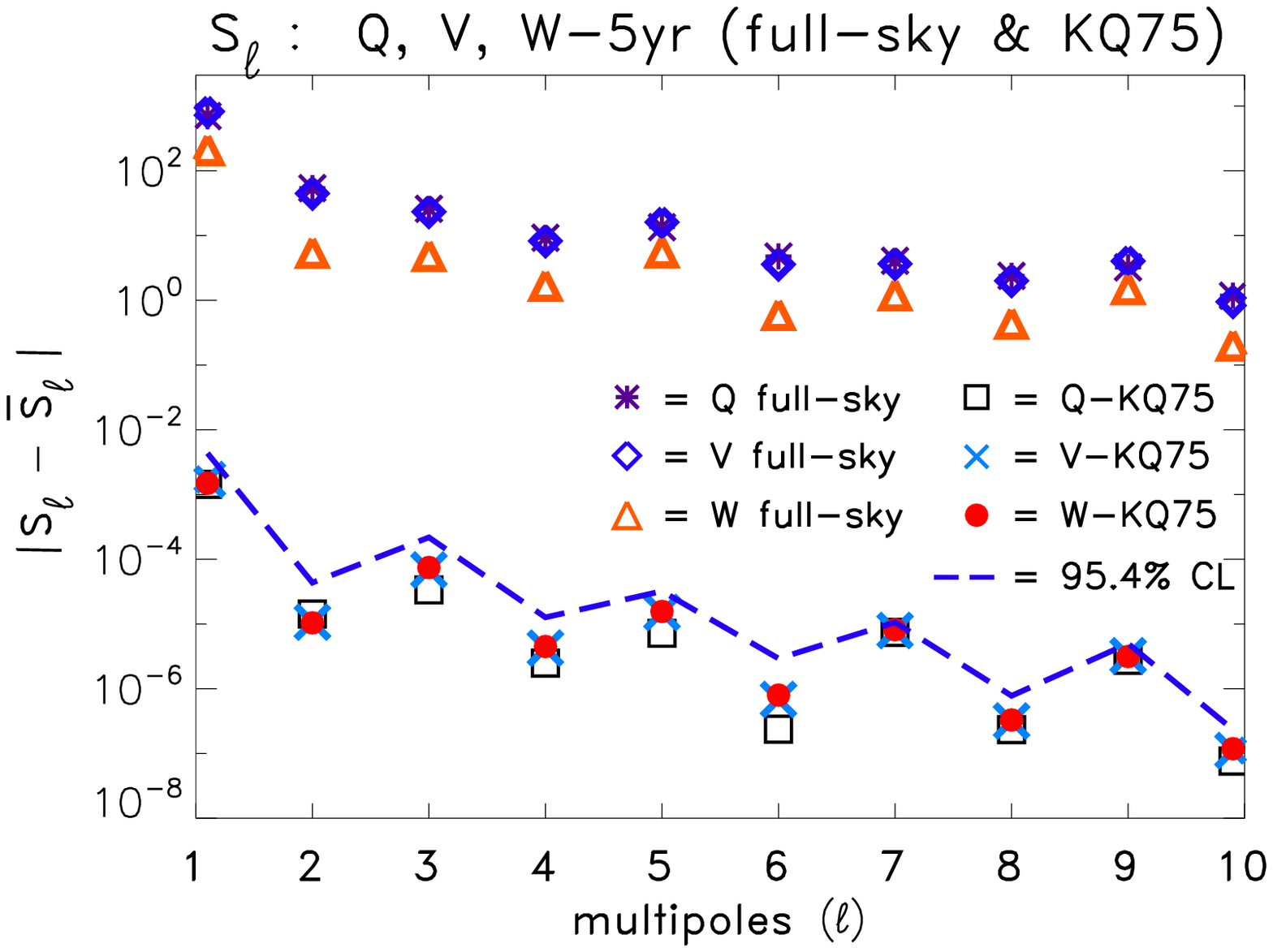,height=7.5cm,width=7.cm}}

\vspace{-7.55cm}
\mbox{\hspace{6.0cm}
\epsfig{file=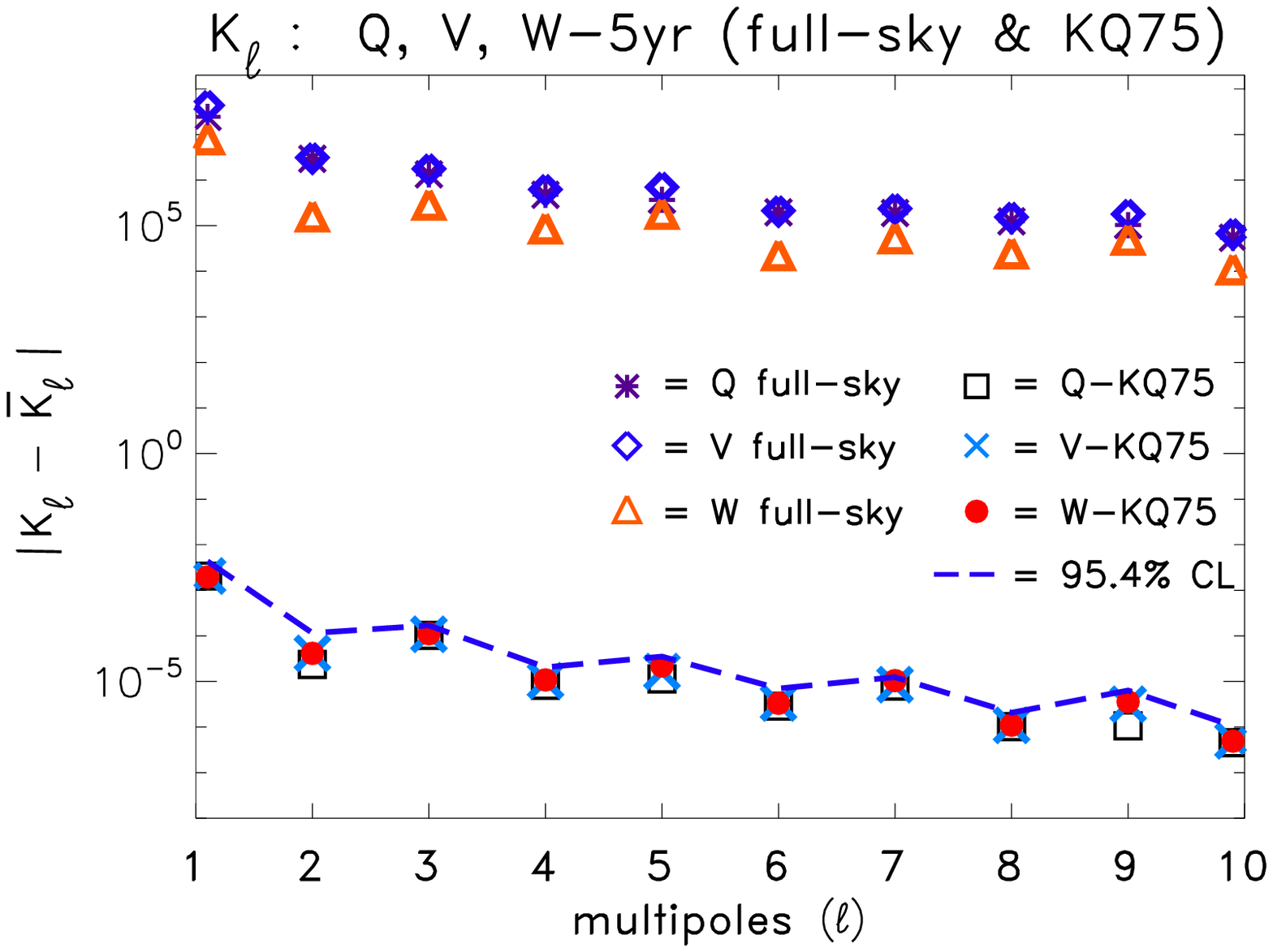,height=7.5cm,width=7.cm}}
\caption{Low $\ell $ \emph{differential} power spectra of skewness
$|S_{\ell} - \overline{S}_{\ell}|$ (left) and kurtosis (right)
$|K_{\ell} - \overline{K}_{\ell}|$
calculated from the WMAP \emph{foreground-reduced} band maps.
The values for the full-sky and \emph{KQ75} masked maps are shown.
The $95.4\%$ confidence level (obtained from $S$ and $K$ maps
calculated from MC Gaussian CMB maps) is indicated by the dashed line.
More details in the text.
\label{fig1}}
\end{figure}

In Fig.~\ref{fig1} we show the angular power spectra of the $S$ maps and $K$ maps
calculated from the WMAP foreground-reduced Q, V, and W CMB band maps by
considering the data in the full-sky and \emph{KQ75} masked maps.
These angular power spectra show that the \emph{KQ75} mask lowers the
non-Gaussianity to a level below  $95.4 \,\%$~CL.

An overall quantitative measure of deviation from Gaussianity
for $\ell=1-10$  is obtained through the $\chi^2$ goodness-of-fit
test by calculating the $\chi^2/\text{dof}$ (where dof stands
for degrees of freedom) of the spectra values $S_{\ell}$ and $K_{\ell}$
calculated from the Q, V, and W band maps as compared to the mean
power spectra $\overline{S}_{\ell}$ and $\overline{K}_{\ell}$
obtained from $S$ and $K$ maps calculated from MC CMB maps.
The results regarding the masked and the full-sky maps are shown in
Table~\ref{table1}. The values obtained in the masked maps analysis
($\chi^2 \lesssim 10$) indicate that these may be seen as consistent
with Gaussianity, while the values  for full-sky foreground-reduced maps
($\chi^2 > 10^{10}$) shows a large deviation from Gaussianity.

\vspace{0.3cm}
\begin{table}
\tbl{Values for the ratio $\chi^2/\text{dof}$ that measure the goodness-of-fit of
the $S_{\ell}$ and $K_{\ell}$ spectra from the WMAP \emph{foreground-reduced}
band maps as compared to the mean spectra $\overline{S}_{\ell}$ and
$\overline{K}_{\ell}$, respectively.
The values for the full-sky and \emph{KQ75} masked maps are shown.}
{\begin{tabular}{@{}lcccccc@{}}
\toprule   
$\chi^2$  $\backslash$ CMB maps & Q [full-sky]  & V [full-sky] & W [full-sky] &
Q [\emph{KQ75}] & V [\emph{KQ75}] & W [\emph{KQ75}] \\
\colrule
$S_{\ell}$  & $9.4 \times 10^{11}$  & $1.2 \times 10^{12}$ & $7.3 \times 10^{10}$ & 3.1 & 4.4 & 3.4 \\
$K_{\ell}$  & $1.2 \times 10^{21}$  & $3.3 \times 10^{21}$ & $1.0 \times 10^{20}$ & 6.3 & 6.5 & 6.0  \\
\botrule     
\end{tabular} \label{table1}}
\end{table}

\section*{Acknowledgments}
This work is supported by Conselho Nacional de Desenvolvimento
Cient\'{\i}fico e Tecnol\'{o}gico (CNPq) - Brasil, under grant No. 472436/2007-4.
\,A.B. and M.J.R. thank CNPq  for their grants.
We acknowledge the use of the Legacy Archive for Microwave Background 
Data Analysis (LAMBDA).\cite{WMAP1} Some of the results in this paper 
were derived using the HEALPix package.\cite{Gorski-et-al-2005}



\end{document}